\documentclass[10pt]{article}
\usepackage{geometry}
\usepackage{xspace}
\usepackage[utf8]{inputenc}
\usepackage{authblk}
\usepackage{amssymb}
\usepackage{amsmath}
\usepackage{amsthm}
\usepackage{threeparttable}
\allowdisplaybreaks[4]
\usepackage{slashed}
\usepackage{graphicx,color,epsfig}
\usepackage{cite}
\usepackage[colorlinks=true,linkcolor=red,citecolor=blue]{hyperref}
\newcommand{\optbar}[1]{\shortstack{{\tiny (\rule[.4ex]{1em}{.1mm})}\\ [-.7ex] $#1$}}

\newcommand{\beq}{\begin{eqnarray}}
\newcommand{\eeq}{\end{eqnarray}}

\newcommand{\KorKbar}{\kern 0.18em\optbar{\kern -0.18em K}{}\xspace}
\def\KorKbarz {\ensuremath{\KorKbar\!^0}\xspace}
\def\KstarIzoptbar {\ensuremath{\KorKbar\!^{*0}(892)}\xspace}

\begin{document}
\title{Branching Fractions and CP Asymmetries of the Quasi-Two-Body Decays in  $B_{s} \to K^0(\overline K^0)K^\pm \pi^\mp$ within PQCD Approach}
\author[1]{Zhi-Tian Zou$\footnote{zouzt@ytu.edu.cn}$}
\author[1,2]{Ying Li$\footnote{liying@ytu.edu.cn}$}
\author[3]{Xin Liu$\footnote{liuxin@jsnu.edu.cn}$}
\affil[1]{\it Department of Physics, Yantai University, Yantai 264005, China}
\affil[2]{\it Center for High Energy Physics, Peking University, Beijing 100871, China}
\affil[3]{\it Department of Physics, Jiangsu Normal University, Xuzhou 221116, China}
\maketitle
\vspace{0.2cm}
\begin{abstract}
Motivated by the first untagged decay-time-integrated amplitude analysis of $B_s \to K_SK^{\mp}\pi^{\pm}$ decays performed by LHCb collaboration, where the decay amplitudes are modeled to contain the resonant contributions from intermediate resonances $K^*(892)$, $K_0^*(1430)$ and $K_2^*(1430)$, we comprehensively investigate the quasi-two-body $B_{s} \to K^0(\overline{\kern -0.2em K}^0 )K^{\pm}\pi^{\mp}$ decays, and calculate the branching fractions and the time-dependent $CP$ asymmetries within the perturbative QCD approach based on the $k_T$ factorization. In the quasi-two-body space region the calculated branching fractions with the considered intermediate resonances are in good agreement with the experimental results of LHCb by adopting proper $K\pi$ pair wave function, describing the interaction between the kaon and pion in the $K\pi$ pair. Furthermore,within the obtained branching fractions of the quasi-two-body decays, we also calculate the branching fractions of corresponding two-body decays, and the results consist with the LHCb measurements and the earlier studies with errors. For these considered decays, since the final states are not flavour-specific, the time-dependent $CP$ could be measured. We then calculate six $CP$-violation observables, which can be tested in the ongoing LHCb experiment.
\end{abstract}
\section{Introduction}
It is well known to us that $B$ meson decays play crucial roles in testing the standard model (SM), understanding the chromodynamics and searching for the possible new physics beyond SM \cite{Cheng:2009xz,Li:2018lxi,Zupan:2019uoi}. In recent years, based on the large data sample, more and more detailed analysis on the $B$ meson three-body hadronic decays have been performed by the BaBar \cite{Aubert:2003mi, Aubert:2005ce, Aubert:2006nu, Aubert:2009av,  Aubert:2007sd, Aubert:2007bs, Aubert:2008bj, Aubert:2009me, Lees:2011nf, BABAR:2011ae, Lees:2012kxa,Aubert:2004cp}, Belle \cite{Abe:2002av, Garmash:2003er, Garmash:2004wa, Garmash:2005rv, Garmash:2006fh, Dalseno:2008wwa, Nakahama:2010nj}, CLEO \cite{Eckhart:2002qr} and LHCb \cite{Aaij:2013sfa, Aaij:2013bla, Aaij:2014iva, Aaij:2016qnm, Aaij:2018rol, Aaij:2019nmr, Aaij:2017zgz, Aaij:2013orb, Aaij:2019hzr, Aaij:2019jaq, Aaij:2013uta, Aaij:2014aaa, Aaij:2015asa, Aaij:2017zpx} collaborations. Due to large phase spaces, some of three-body $B$ decays have the branching fractions as large as $10^{-5}$, which can be measured precisely with small uncertainties. In the theoretical side, three-body non-leptonic $B$ decays are interesting for several phenomenological applications, such as the study of $CP$ violation and the extraction of the CKM angles $\alpha$ \cite{Snyder:1993mx} and $\gamma$ \cite{Grossman:2002aq}. The abundant measurements and phenomenology in three-body decays attract considerable theoretical interests in understanding three-body hadronic $B$ meson decays, as a result, studies of various three-body hadronic $B$ decays have been accomplished in different frameworks, such as the symmetry principles \cite{Gronau:2005ax, Engelhard:2005hu, Imbeault:2011jz, Bhattacharya:2013boa, He:2014xha}, the QCD factorization (QCDF) \cite{ElBennich:2009da, Krankl:2015fha, Cheng:2002qu, Cheng:2007si, Cheng:2016shb, Cheng:2014uga, Cheng:2013dua, Li:2014oca}, the PQCD approach \cite{Wang:2014ira, Wang:2015uea, Wang:2016rlo, Li:2016tpn, Wang:2017hao, Ma:2019sjo, Li:2018qrm, Li:2018lbd, Ma:2017aie, Li:2017obb, Ma:2017kec, Li:2017mao, Ma:2016csn, Li:2015tja, Ma:2017idu, Rui:2017bgg, Rui:2017fje, Rui:2018hls, Rui:2019yxx, Li:2019pzx, Li:2019hnt, Rui:2017hks, Xing:2019xti, Cui:2019khu, Rui:2018mxc, Wang:2020saq,Zou:2020atb, Wang:2020plx}, and other theoretical methods\cite{Zhang:2013oqa, Wang:2015ula, Qi:2018syl, ElBennich:2006yi,Cheng:2019tgh}.

In contrast to the two-body decays where the momenta of final states are fixed, the momenta of final states of three-body vary in certain ranges. Strong dynamics contained in three-body hadronic $B$ meson decays is much more complicated than that in two-body cases, because of entangled nonresonant and resonant contributions, and significant final-state interactions. In particular, the Dalitz plot analysis has been adopted for studying the three-body decays. The Dalitz plot can be divided into different regions with characteristic kinematics. In the central region where three daughters with large energy fly apart from each other in the $B$ rest frame at about $120^{\circ}$ angle, the contribution is nonresonant and both power- and $\alpha_s$-suppressed with respect to the amplitude at the edge \cite{Virto:2016fbw}. The corners of the Dalitz plot correspond to the cases in which there is one particle at almost rest, while the other two particles fly back-to-back. At the edge of the Dalitz plot where the three mesons are quasi aligned, two mesons move collinearly and recoil against the third meson.  We will denote such processes as $B\to (M_1M_2)M_3$ where the mesons of the $ M_1M_2 $ pair, move, more or less, in the same direction. The bachelor particle $M_3$ moves in the opposite direction. In such condition, the three-body interactions are expected to suppressed. Thereby, it is reasonable to assume the validity of factorization for this quasi–two-body $B$ decay \cite{Beneke:2007zz} where we assume that the quasi-two-body final state $M_1M_2$ pair originates from a quark-antiquark state. In some certain energy regions, some resonant structures can be seen. For the $K\pi$ pair, one clearly observes a vector $K^*$, a scalar $K_0^*(1430)$ and a tensor $K_2^*(1430)$ in $B\to K\pi\pi$ decays \cite{Aubert:2005ce,Garmash:2006fh, Aubert:2007bs, Aubert:2008bj}.

In the decays of $B$ mesons, the decay modes induced by the flavor-changing neutral-current $b\to s$ are of interest to search for the new source of CP asymmetry beyond the Cabibbo-Kobayashi-Maskawa (CKM) mechanism and to probe the new physics beyond SM, because the new particles could affect the observables by entering the loops. Along this line, the analysis of $B_s \to K_S K^\pm \pi^\mp$ decays have been preformed by LHCb collaboration \cite{Aaij:2013uta, Aaij:2014aaa, Aaij:2015asa, Aaij:2017zpx}, and the corresponding quasi-two-body decays with respect to the resonances $K^{*}$ and $K_0^*$ have also been explored. For these decays, the $K_S K^+ \pi^-$ and $K_S K^- \pi^+$ final states are not flavour-special and as such both $B_s^0$ and ${\overline B}_s^0$ decays can contribute to each, with the corresponding amplitudes expected to be comparable in magnitude. The resonant and nonresonant contributions can provide different sources of strong phases, so large interference effects and potentially large CP-violation effects are possible. Very recently in ref.\cite{Aaij:2019nmr}, the LHCb collaboration released the first untagged decay-time-integrated amplitude analysis to these $B_{s} \to K_S K^{\pm}\pi^{\mp}$ decays with the $K^*(892)$, $K_0^*(1430)$ and $K_2^*(1430)$ intermediate resonances using a sample corresponding to $3.0~{\rm fb}^{-1}$ of $pp$ collision data, and the observed branching ratios of quasi-two-body decays were also reported. Motivated by above results, we shall investigate the four $B_{s} (\overline{B}_s^0)\to K_S K^{\pm}\pi^{\mp}$ decays in the regions relate to the $K^*(892)$, $K_0^*(1430)$ and $K_2^*(1430)$ resonances within the PQCD approach. Besides the branching fractions of these quasi-two-body decays, we will calculate the direct CP violations and the time-dependent $CP$-violations in detail, which could be measured in the ongoing LHCb experiment.

As aforementioned, in the quasi-two-body $B\to (M_1M_2)M_3$ decays, the two mesons $M_1M_2$ move collinearly fast, and the bachelor meson $M_3$ is also energetic and recoil against the meson pair in the $B$ meson rest frame. The interaction between the meson pair and the bachelor meson is viewed as to be power suppressed. At the quark level this configuration involves the hadronization of two energetic collinear quarks, produced from the $b$ quark decay, into the two collimated hadrons. In this picture, the factorization formula for the $B\to (M_1M_2)M_3$ decay is then expressed to be the convolution as \cite{Chen:2002th,Chen:2004az}
\begin{eqnarray} \label{amp}
\mathcal{A}\sim \Phi_B \otimes {\cal H}\otimes \Phi_{M_1M_2}\otimes \Phi_{M_3},
\end{eqnarray}
where $\cal H$ is the hard kernel, $\Phi_B$ and $\Phi_{M_3}$ being the universal wave functions of the $B$ meson and the bachelor meson, respectively. The resonant and nonresonant interactions between the two moving collinearly mesons are all included in the two-meson wave function $\Phi_{M_1M_2}$. In PQCD approach based on $k_T$ factorization \cite{Keum:2000ph, Lu:2000em, Chang:1996dw} these decays are governed by the transition with a hard gluon exchanging between the spectator quark and the quark involved in the four-quark operator, making the hard kernel become the six-quark interaction rather than the traditional four-quark interaction. The hard kernel $\cal H$ can be calculated perturbatively away from the endpoint singularity, as the intrinsic transverse momenta of the inner quarks are retained. Besides the hard gluon exchange with the spectator quark, the soft gluon exchanges between quark lines give out the double logarithms from the overlap of collinear and soft divergence. The resummation of these double logarithms leads to a Sudakov form factor, which could suppresses the long distance contributions. More details about the PQCD can be found in refs.\cite{Li:2003yj,Li:2014rwa}.

The rest of the paper is organized as follows. In Sec.\ref{sec:function}, we will introduce the decay formalism including the weak decay Hamiltonian, the two-meson wave functions and the mixing in $B_s-\overline{B}_s$ system.  The total decay amplitudes with the wilson coefficients, CKM matrix elements and the amplitudes of four-quark operators will be presented in Sec.~\ref{sec:function}. In Sec.~\ref{sec:result}, we will address the numerical results including the branching fractions and the time-dependent $CP$-violation observables. Lastly, we summarize this work in Sec.~\ref{summary}.
\section{Framework and Amplitudes}\label{sec:function}
In this section, we will start with the effective weak Hamiltonian for the $b\to s$ transitions, which are
given by \cite{Buchalla:1995vs}
\begin{eqnarray}
\mathcal{H}_{eff}=\frac{G_F}{\sqrt{2}}\{V^*_{ub}V_{us}(C_1O_1+C_2O_2)-V^*_{tb}V_{ts}\sum_{i=3}^{10}C_i O_i\},
\end{eqnarray}
where $V_{ub(s)}$ and $V_{tb(s)}$ are the CKM matrix elements. The explicit expressions of the local four-quark operators $O_i$ ($i = 1, ..., 10$) and the corresponding wilson coefficients $C_i$ at different scales have been given in Ref.\cite{Buchalla:1995vs}. Note that $O_1$ and $O_2$ are tree operators and others $O_{3-10}$ are penguin ones. Noted that the interference between the contributions from tree operators and the ones from the penguins leads to the $CP$ asymmetry in SM.

In what follows, for the sake of brevity, we shall take the decay ${\overline B}_s^0 \to (K^0\pi^+)K^-$ as an example for illustration.
From eq.(\ref{amp}), in order to calculate the decay amplitude of this decay, the wave functions of $\Phi_{B_s}$, $\Phi_{K}$ and $\Phi_{K\pi}$ are needed. It is true that in PQCD approach the wave functions are the most important inputs which affect remarkably the predictions and are the main sources of the theoretical uncertainties. For the $B_s$ meson and $K$ meson, the wave functions have been well discussed and established in the charm/charmless two-body decays \cite{Liu:2019ymi, Zou:2016yhb, Qin:2014xta, Yu:2013pua, Zou:2012sx, Li:2015xna, Wang:2017hxe, Zhou:2015jba, Yu:2005rh, Li:2004ep, Colangelo:2010wg, Colangelo:2010bg, Wang:2016wpc}. For the $K\pi$-pair, a wave function $\Phi_{K\pi}$  describes the hadronization of two collinear quarks, together with other quarks popped out of the vacuum, into two collimated mesons. Now, although the exact form of the wave function based on QCD-inspired approach is absent, there are many phenomenological attempts based on the experimental measurements, and the involved parameters can be constrained. In this work, we will also adopt the analytic forms that have been constrained from $B\to K\pi\pi$ \cite{Li:2018qrm}, $B\to KK\pi$\cite{Wang:2020saq}and $B\to\psi K\pi$ \cite{Li:2019hnt}. In the current work, we will follow the analysis of LHCb \cite{Aaij:2019nmr}, and account for the $S$-wave, $P$-wave and $D$-wave resonances corresponding to $K_0^*(1430)$, $K^*(892)$ and $K_2^*(1430)$, respectively.

We first present the $S$-wave $K\pi$-pair wave function $\Phi_S$ as\cite{Li:2019hnt}
\begin{eqnarray}
\Phi_S=\frac{1}{2\sqrt{N_c}}\Big[P\!\!\!\!\slash\phi_S(z,\zeta,\omega)+\omega\phi_S^s(z,\zeta,\omega)+\omega(n\mkern-10.5mu/ v\mkern-10.5mu/-1)\phi_S^t(z,\zeta,\omega)\Big],
\end{eqnarray}
where $P$ is the momentum of the $K\pi$-pair in the ${\overline B}_s^0$ meson rest framework, and $n=(1, 0, 0_T)$ and $v=(0, 1, 0_T)$ are the dimensionless vectors. $\phi_S$ is the twist-2 light-cone distribution amplitude (DA), and $\phi_S^{s,t}$ are the twist-3 ones. In the DAs, $z$ is the momentum fraction of the spectator quark, and $\zeta$ is the momentum fraction of the $K$ in the $K\pi$-pair. $\omega$ is the invariant mass of the $K\pi$-pair satisfying $\omega^2=P^2$. The light-cone DAs can be expanded in terms of the Gegenbauer polynomials such as $C_{1,3}^{3/2}$ with corresponding Gegenbauer moments as \cite{Li:2019hnt}:
\begin{eqnarray}
\phi_S(z,\zeta,\omega)&=&\frac{6}{2\sqrt{2N_c}}F_S(\omega)z(1-z)\left[\frac{1}{\mu_S}+B_1C_1^{3/2}(1-2z)+B_3C_3^{3/2}(1-2z)\right],\\
\phi_S^s(z,\zeta,\omega)&=&\frac{1}{2\sqrt{2N_c}}F_S(\omega),\\
\phi_S^t(z,\zeta,\omega)&=&\frac{1}{2\sqrt{2N_c}}F_S(\omega)(1-2z),
\end{eqnarray}
with $\mu_S=\omega/(m_2-m_1)$ where $m_{1,2}$ are the masses of the running current quarks in the resonance $K_0^*(1430)$.  For $B_1$ and $B_3$, unlike in the refs.\cite{Wang:2020saq} where the authors have adopt the same values as ones of the $K_0^*(1430)$ and large uncertainties were taken, we here determined these values to be $B_1=-0.4$ and $B_3=-0.8$ within the current experimental data \cite{Aaij:2019nmr}, with which we then predict the six $CP$-violation observables later. $F_S(\omega)$ is the time-like form factor, and in particular it is often parameterized by the relativistic Breit-Wigner (RBW) model. However, the RBW function is a good model for narrow resonances that are well separated from any other resonant or nonresonant contribution of the same spin. This approach is known to break down in the $K\pi$ $S$-wave because the $K_0^*(1430)$  resonance interferes strongly with a slowly varying nonresonant term \cite{Meadows:2007jm}. In view of this, the LASS line shape \cite{Aston:1987ir, Back:2017zqt} is developed to evaluate the combined amplitude, the expression of which is given by
\begin{eqnarray}
F_S(\omega)=\frac{\omega}{|p_1|(\cot \delta -i)}+e^{2i\delta}\frac{m_0\Gamma_0\frac{m_0}{|p_0|}}
{m_0^2-\omega^2-im_0\Gamma_0\frac{|p_1|}{\omega}\frac{m_0}{|p_0|}},
\label{lass}
\end{eqnarray}
with $\cot \delta =\frac{1}{a|p_1|}+\frac{r|p_1|}{2}$. The $m_0$ and $\Gamma_0$ are the pole mass and width of the cresponding resonance $K_0^*(1430)$, respectively. $|p_1|$ is the magnitude of the momentum of one of daughter of resonance in the center-of-mass frame of the meson pair, the value of which is $|p_0|$ when the invariant mass $\omega$ equals to $m_0$. The shape parameters $a=1.95~{\rm GeV}^{-1}$ and $r=1.95~{\rm GeV}^{-1}$ \cite{web} are the scattering length and the effective range, respectively. It is noted that the first term represents a background i.e. nonresonant contribution, while the second is the resonant contribution. In this work, we shall adopt the LASS line shape, as done in the LHCb experiment \cite{Aaij:2019nmr}.

The wave function of $P$-wave $K\pi$ pair is very similar to that of the vector meson. Due to the angular momentum conservation, only the longitudinal wave function contribute to the decay modes we concerned, and it is written as \cite{Li:2019hnt}
\begin{eqnarray}
\Phi_P=\frac{1}{\sqrt{2N_c}}\left[P\mkern-10.5mu/\phi_P(z,\zeta,\omega)+\omega\phi_P^s(z,\zeta,\omega)+\frac{P\mkern-10.5mu/_1 P\mkern-10.5mu/_2-P\mkern-10.5mu/_2P\mkern-10.5mu/_1}{\omega(2\zeta-1)}\phi_P^t(z,\zeta,\omega)\right],
\label{Pwave}
\end{eqnarray}
with the twist-2, 3 light-cone distribution amplitudes
\begin{eqnarray}
\phi_P(z,\zeta,\omega)&=&\frac{3F_P^{\parallel}(\omega)}{\sqrt{2N_C}}z(1-z)\left[1+a_1C_1^{3/2}(t)+a_2C_2^{3/2}(t)\right](2\zeta-1-\xi),\\
\phi_P^s(z,\zeta,\omega)&=&\frac{3F_P^{\perp}(\omega)}{2\sqrt{2N_C}}\Big[t(1+a_s t)-a_s 2z(1-z)\Big](2\zeta-1),\\
\phi_P^t(z,\zeta,\omega)&=&\frac{3F_P^{\perp}(\omega)}{2\sqrt{2N_C}}t\Big[t+a_t(3t^2-1)\Big](2\zeta-1),
\end{eqnarray}
and $t=1-2z$. The parameter $\xi$ is defined as
\begin{eqnarray}
\xi=\frac{m_K^2-m_{\pi}^2}{\omega^2},
\end{eqnarray}
which reflects the mass difference between kaon and pion in the $K\pi$ pair. The Gegenbauer moments $a_i$ ($i=1,2,s,t$) have been determined in the refs.\cite{Li:2019hnt,Li:2019pzx} and are taken as
\begin{eqnarray}
a_1=0.2,\;\;a_2=0.5,\,\;\;a_s=-0.2,\;\;a_t=0.2.
\end{eqnarray}
For the $P$-wave time-like form factor $F_P^{\parallel}$, we adopt the RBW model \cite{Back:2017zqt} and present the function as
\begin{eqnarray}
F_P^{\parallel}(\omega)=\frac{c m_{0}^2}{m_{0}^2-\omega^2-im_{0}\Gamma(\omega)},\label{rbw}
\end{eqnarray}
where $m_0$ is the nominal mass of the resonance. For a resonance with spin-$L$, the mass-dependent width $\Gamma(\omega)$ can be expressed as \cite{Lees:2012kxa,Back:2017zqt}
\begin{eqnarray}
\Gamma(\omega)=\Gamma_0\left(\frac{|p_1|}{|p_0|}\right)^{2L+1}\left(\frac{m_0}{\omega}\right)\mathrm{X}_L^2(r| p_1 |),
\end{eqnarray}
where the definitions of $|p_0|$ and $| p_1|$ are same as ones in eq.(\ref{lass}). The Blatt-Weillkopf barrier factor $\mathrm{X}_L$ \cite{Blatt:1952ije} are angular momentum dependent and are given by
\begin{eqnarray}
L=0,\;\;\mathrm{X}(a)=1,\\
L=1,\;\;\mathrm{X}(a)=\sqrt{\frac{1+a_0^2}{1+a^2}},\\
L=2,\;\;\mathrm{X}(a)=\sqrt{\frac{a_0^4+3a_0^2+9}{a^4+3a^2+9}},
\end{eqnarray}
where the $a_0$ is the value of $a$ at the pole mass of the resonance. The effective meson radius $r$ is taken to be $4~{\rm GeV}^{-1}~\approx~0.8~{\rm fm}$ \cite{Aubert:2005ce} for each resonance. Note that the value of $r$ does not affect the predictions remarkably. Here we follow the definition and the determination of the ref.\cite{Li:2019hnt} and adopt $c=0.72$. We also point out that the coefficient $c$ can be absorbed in the value of Gegenbauer moments when only considering singlet resonance $K^*(892)$.  As for the transverse time-like form factor $F_P^{\perp}$,  we follow ref.\cite{Wang:2016rlo} and use the relation as
\begin{eqnarray}
\frac{F_P^{\perp}}{F_P^{\parallel}}\approx \frac{f_V^T}{f_V},
\label{fprep}
\end{eqnarray}
where the $f_V^{(T)}$ is the vector (tensor) decay constant of the $P$-wave resonance $K^*(892)$.

For the $D$-wave $K\pi$-pair with the spin $L=2$, the helicity $\lambda=\pm2$ components do not contribute because of the angular momentum conservation. So, the form of the $D$-wave $K\pi$-pair is almost similar to that of $P$-wave pair shown in eq.(\ref{Pwave}) with the different distribution amplitudes $\phi_T$, $\phi_T^{s}$ and $\phi_T^{t}$ as
\begin{eqnarray}
\phi_T(z,\zeta,\omega)&=&\sqrt{\frac{2}{3}}\frac{6F_D^{\parallel}(\omega)}{2\sqrt{2N_c}}z(1-z)\left[3a_D(2z-1)\right]P_2(2\zeta-1),\\
\phi_T^s(z,\zeta,\omega)&=&\sqrt{\frac{2}{3}}\frac{-9F_D^{\perp}(\omega)}{4\sqrt{2N_c}}\left[a_D(1-6z+6z^2)\right]P_2(2\zeta-1),\\
\phi_T^t(z,\zeta,\omega)&=&\sqrt{\frac{2}{3}}\frac{9F_D^{\perp}(\omega)}{4\sqrt{2N_c}}\left[a_D(1-6z+6z^2)(2z-1)\right]P_2(2\zeta-1),
\end{eqnarray}
with the Gegenbauer moment $a_D=0.5$ and the Legendre polynomial $P_2(x)=\frac{1}{2}(3x^2-1)$. $F_D^{\parallel}$ and $F_D^{\perp}$ are the $D$-wave time-like form factors, and they are described by the RBW model in eq.(\ref{rbw}) and eq.(\ref{fprep}) with the parameters of the corresponding resonance $K_2^*(1430)$ meson \cite{Tanabashi:2018oca}.

For simplicity, we work in the rest frame of the ${\overline B}_s^0$ meson. For the mode ${\overline B}_s^0 \to (K^0\pi^+)K^-$, in the light-cone coordinates the ${\overline B}_s^0$ momentum $p_B$, the $P\pi$ pair momentum $P$ and the bachelor $K^-$ momentum $p_3$ can be written as \cite{Wang:2016rlo, Li:2016tpn}
\begin{eqnarray}
p_B=\frac{m_{B_s}}{\sqrt{2}}(1,1,\vec{0}_{\rm T}),\,\,\,
P=\frac{m_{B_s}}{\sqrt2}(1,\eta^2,\vec{0}_{\rm T}),\,\,\,
p_3=\frac{m_{B_s}}{\sqrt2}(0,1-\eta^2,\vec{0}_{\rm T}),
\end{eqnarray}
with $m_{B_s}$ being the ${\overline B}_s^0$ meson mass and $\eta=w /m_{B_s}$. The momenta of the light spectator quark $\bar s$ in ${\overline B}_s^0$ and $P\pi$ pair are denoted as $k_B$ and $k_P$, and the momentum of the light quark in the bachelor $K^-$ is $k_3$, and they are given by
\begin{eqnarray}
k_B=(0 ,\frac{m_{B_s}}{\sqrt{2}}x,\vec{\rm k}_{\rm {1T}}),\,\,\,
k_P=\left(\frac{m_{B_s}}{\sqrt2}z,0, \vec{\rm k}_{\rm{PT}}\right),\,\,\,
k_3=\left(0,\frac{m_{B_s}}{\sqrt2}(1-\eta^2)x_3, {\vec{\rm k}}_{\rm{3T}}\right),
\end{eqnarray}
where $x$, $z$ and $x_3$ are the momentum fractions. In the $K\pi$ pair, the momenta $p_1$ and $p_2$ for the kaon and pion have the components as
\begin{eqnarray}
p^+_1=\zeta\frac{m_{B_s}}{\sqrt2},           \quad
p^-_1=(1-\zeta)\eta^2 \frac{m_{B_s}}{\sqrt2}, \quad
p^+_2=(1-\zeta)\frac{m_{B_s}}{\sqrt2},       \quad
p^-_2=\zeta\eta^2\frac{m_{B_s}}{\sqrt2},\label{def-pp4}
\end{eqnarray}
with $\zeta$ varying in $[0,1]$.

\begin{figure}[!htb]
\begin{center}
\includegraphics[scale=1.0]{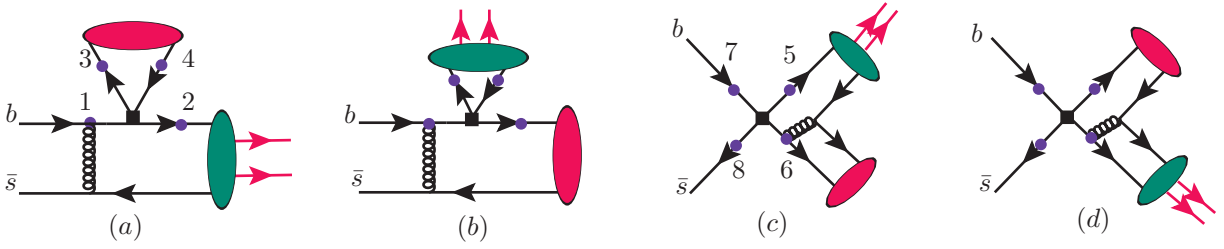}
\caption{Typical Feynman diagrams for the $\overline B_s^0\to KK\pi$ in PQCD, where the black squares stand for the weak vertices, and large (purple) spots on the quark lines denote possible attachments of hard gluons. The green ellipses represent $K\pi$-pair and the red ones are the light bachelor mesons.}\label{feynman}
\end{center}
\end{figure}

Based on the aforementioned two-meson wave functions introduced and dynamical conventions, we can calculate each diagrams shown in Fig.~\ref{feynman} in PQCD approach. The amplitudes of ${\overline B}_s^0 \to (K^0\pi^+)K^- $, $({\overline K}^0\pi^-)K^+$, $(K^-\pi^+)K^0$ and $(K^+\pi^-){\overline K}^0$ are given as:
\begin{multline}
 \mathcal{A}({\overline B}_s^0 \to (K^0\pi^+)K^-)\\
=\frac{G_F}{\sqrt{2}}\Bigg\{V_{ub}V_{us}^*
\Bigg(\mathcal{F}_{K^0\pi^+}^{LL}\Big[\frac{1}{3}C_1+C_2\Big]+\mathcal{M}_{K^0\pi^+}^{LL}\Big[C_1\Big]
+\mathcal{A}_{K^0\pi^+,K^-}^{LL}\Big[C_1+\frac{1}{3}C_2\Big]+\mathcal{W}_{K^0\pi^+,K^-}^{LL}\Big[C_2\Big]\Bigg)\\
-V_{tb}V_{ts}^*\Bigg(\mathcal{F}_{K^0\pi^+}^{LL}\Big[\frac{1}{3}C_3+C_4+\frac{1}{3}C_9+C_{10}\Big]
+\mathcal{F}_{K^0\pi^+}^{SP}\Big[\frac{1}{3}C_5+C_6+\frac{1}{3}C_7+C_8\Big]\\
+\mathcal{M}_{K^0\pi^+}^{LL}\Big[C_3+C_9\Big]+\mathcal{M}_{K^0\pi^+}^{LR}\Big[C_5+C_7\Big]
+\mathcal{A}_{K^0\pi^+,K^-}^{LL}\Big[C_3+\frac{1}{3}C_4+C_9+\frac{1}{3}C_{10}\Big]\\
+\mathcal{A}_{K^0\pi^+,K^-}^{LR}\Big[C_5+\frac{1}{3}C_6+C_7+\frac{1}{3}C_{8}\Big]+\mathcal{W}_{K^0\pi^+,K^-}^{LL}\Big[C_4+C_{10}\Big]
+\mathcal{W}_{K^0\pi^+,K^-}^{SP}\Big[C_6+C_{8}\Big]\\
+\mathcal{A}_{K^-,K^0\pi^+}^{LL}\Big[\frac{4}{3}C_3+\frac{4}{3}C_4-\frac{2}{3}C_9-\frac{2}{3}C_{10}\Big]
+\mathcal{A}_{K^-,K^0\pi^+}^{LR}\Big[C_5+\frac{1}{3}C_6-\frac{1}{2}C_7-\frac{1}{6}C_{8}\Big]\\
+\mathcal{A}_{K^-,K^0\pi^+}^{SP}\Big[\frac{1}{3}C_5+C_6-\frac{1}{6}C_7-\frac{1}{2}C_{8}\Big]
+\mathcal{W}_{K^-,K^0\pi^+}^{LL}\Big[C_3+C_4-\frac{1}{2}C_9-\frac{1}{2}C_{10}\Big]\\
+\mathcal{W}_{K^-,K^0\pi^+}^{LR}\Big[C_5-\frac{1}{2}C_{7}\Big]
+\mathcal{W}_{K^-,K^0\pi^+}^{SP}\Big[C_6-\frac{1}{2}C_{8}\Big]
\Bigg) \Bigg\}, \label{af}
\end{multline}
\begin{multline}
 \mathcal{A}({\overline B}_s^0 \to ({\overline K}^0\pi^-)K^+)\\
=\frac{G_F}{\sqrt{2}}\Bigg\{V_{ub}V_{us}^*\Bigg(\mathcal{F}_{K^+}^{LL}\Big[\frac{1}{3}C_1+C_2\Big]+\mathcal{M}_{K^+}^{LL}\Big[C_1\Big]
+\mathcal{A}_{K^+,{\overline K}^0\pi^-}^{LL}\Big[C_1+\frac{1}{3}C_2\Big]+\mathcal{W}_{K^+,{\overline K}^0\pi^-}^{LL}\Big[C_2\Big]\Bigg)\\
-V_{tb}V_{ts}^*\Bigg(\mathcal{F}_{K^+}^{LL}\Big[\frac{1}{3}C_3+C_4+\frac{1}{3}C_9+C_{10}\Big]
+\mathcal{F}_{K^+}^{SP}\Big[\frac{1}{3}C_5+C_6+\frac{1}{3}C_7+C_8\Big]\\
+\mathcal{M}_{K^+}^{LL}\Big[C_3+C_9\Big]+\mathcal{M}_{K^+}^{LR}\Big[C_5+C_7\Big]
+\mathcal{A}_{K^+,{\overline K}^0\pi^-}^{LL}\Big[C_3+\frac{1}{3}C_4+C_9+\frac{1}{3}C_{10}\Big]\\
+\mathcal{A}_{K^+,{\overline K}^0\pi^-}^{LR}\Big[C_5+\frac{1}{3}C_6+C_7+\frac{1}{3}C_{8}\Big]
+\mathcal{W}_{K^+,{\overline K}^0\pi^-}^{LL}\Big[C_4+C_{10}\Big]
+\mathcal{W}_{K^+,{\overline K}^0\pi^-}^{SP}\Big[C_6+C_{8}\Big]\\
+\mathcal{A}_{{\overline K}^0\pi^-,K^+}^{LL}\Big[\frac{4}{3}C_3+\frac{4}{3}C_4-\frac{2}{3}C_9-\frac{2}{3}C_{10}\Big]
+\mathcal{A}_{{\overline K}^0\pi^-,K^+}^{LR}\Big[C_5+\frac{1}{3}C_6-\frac{1}{2}C_7-\frac{1}{6}C_{8}\Big]\\
+\mathcal{A}_{{\overline K}^0\pi^-,K^+}^{SP}\Big[\frac{1}{3}C_5+C_6-\frac{1}{6}C_7-\frac{1}{2}C_{8}\Big]
+\mathcal{W}_{{\overline K}^0\pi^-,K^+}^{LL}\Big[C_3+C_4-\frac{1}{2}C_9-\frac{1}{2}C_{10}\Big]\\
+\mathcal{W}_{{\overline K}^0\pi^-,K^+}^{LR}\Big[C_5-\frac{1}{2}C_{7}\Big]
+\mathcal{W}_{{\overline K}^0\pi^-,K^+}^{SP}\Big[C_6-\frac{1}{2}C_{8}\Big]
\Bigg) \Bigg\},\label{afbar}
\end{multline}
\begin{multline}
 \mathcal{A}({\overline B}_s^0 \to (K^+\pi^-){\overline K}^0)\\
=-\frac{G_F}{\sqrt{2}}V_{tb}V_{ts}^*\Bigg(
 \mathcal{F}_{K^+\pi^-}^{LL}\Big[\frac{1}{3}C_3+C_4-\frac{1}{6}C_9-\frac{1}{2}C_{10}\Big]
+\mathcal{F}_{K^+\pi^-}^{SP}\Big[\frac{1}{3}C_5+C_6-\frac{1}{6}C_7-\frac{1}{2}C_8\Big]\\
+\mathcal{M}_{K^+\pi^-}^{LL}\Big[C_3-\frac{1}{2}C_9\Big]
+\mathcal{M}_{K^+\pi^-}^{LR}\Big[C_5-\frac{1}{2}C_7\Big]
+\mathcal{A}_{{\overline K}^0,K^+\pi^-}^{LL}\Big[\frac{4}{3}C_3+\frac{4}{3}C_4-\frac{2}{3}C_9-\frac{2}{3}C_{10}\Big]\\
+\mathcal{A}_{{\overline K}^0,K^+\pi^-}^{LR}\Big[C_5+\frac{1}{3}C_6-\frac{1}{2}C_7-\frac{1}{6}C_{8}\Big]
+\mathcal{A}_{{\overline K}^0,K^+\pi^-}^{SP}\Big[\frac{1}{3}C_5+C_6-\frac{1}{6}C_7-\frac{1}{2}C_{8}\Big]\\
+\mathcal{W}_{{\overline K}^0,K^+\pi^-}^{LL}\Big[C_3+C_4-\frac{1}{2}C_9-\frac{1}{2}C_{10}\Big]
+\mathcal{W}_{{\overline K}^0,K^+\pi^-}^{LR}\Big[C_5-\frac{1}{2}C_{7}\Big]
+\mathcal{W}_{{\overline K}^0,K^+\pi^-}^{SP}\Big[C_6-\frac{1}{2}C_{8}\Big]\\
+\mathcal{A}_{K^+\pi^-,{\overline K}^0}^{LL}\Big[C_3+\frac{1}{3}C_4-\frac{1}{2}C_9-\frac{1}{6}C_{10}\Big]
+\mathcal{A}_{K^+\pi^-,{\overline K}^0}^{LR}\Big[C_5+\frac{1}{3}C_6-\frac{1}{2}C_7-\frac{1}{6}C_{8}\Big]\\
+\mathcal{W}_{K^+\pi^-,{\overline K}^0}^{LL}\Big[C_4-\frac{1}{2}C_{10}\Big]
+\mathcal{W}_{K^+\pi^-,{\overline K}^0}^{SP}\Big[C_6-\frac{1}{2}C_{8}\Big]
\Bigg),
\end{multline}
\begin{multline}
 \mathcal{A}({\overline B}_s^0 \to (K^-\pi^+)K^0)\\
=-\frac{G_F}{\sqrt{2}}V_{tb}V_{ts}^*\Bigg(
 \mathcal{F}_{K^0}^{LL}\Big[\frac{1}{3}C_3+C_4-\frac{1}{6}C_9-\frac{1}{2}C_{10}\Big]
+\mathcal{F}_{K^0}^{SP}\Big[\frac{1}{3}C_5+C_6-\frac{1}{6}C_7-\frac{1}{2}C_8\Big]\\
+\mathcal{M}_{K^0}^{LL}\Big[C_3-\frac{1}{2}C_9\Big]
+\mathcal{M}_{K^0}^{LR}\Big[C_5-\frac{1}{2}C_7\Big]
+\mathcal{A}_{K^-\pi^+,K^0}^{LL}\Big[\frac{4}{3}C_3+\frac{4}{3}C_4-\frac{2}{3}C_9-\frac{2}{3}C_{10}\Big]\\
+\mathcal{A}_{K^-\pi^+,K^0}^{LR}\Big[C_5+\frac{1}{3}C_6-\frac{1}{2}C_7-\frac{1}{6}C_{8}\Big]
+\mathcal{A}_{K^-\pi^+,K^0}^{SP}\Big[\frac{1}{3}C_5+C_6-\frac{1}{6}C_7-\frac{1}{2}C_{8}\Big]\\
+\mathcal{W}_{K^-\pi^+,K^0}^{LL}\Big[C_3+C_4-\frac{1}{2}C_9-\frac{1}{2}C_{10}\Big]
+\mathcal{W}_{K^-\pi^+,K^0}^{LR}\Big[C_5-\frac{1}{2}C_{7}\Big]
+\mathcal{W}_{K^-\pi^+,K^0}^{SP}\Big[C_6-\frac{1}{2}C_{8}\Big]\\
+\mathcal{A}_{K^0,K^-\pi^+}^{LL}\Big[C_3+\frac{1}{3}C_4-\frac{1}{2}C_9-\frac{1}{6}C_{10}\Big]
+\mathcal{A}_{K^0,K^-\pi^+}^{LR}\Big[C_5+\frac{1}{3}C_6-\frac{1}{2}C_7-\frac{1}{6}C_{8}\Big]\\
+\mathcal{W}_{K^0,K^-\pi^+}^{LL}\Big[C_4-\frac{1}{2}C_{10}\Big]
+\mathcal{W}_{K^0,K^-\pi^+}^{SP}\Big[C_6-\frac{1}{2}C_{8}\Big]
\Bigg).
\end{multline}

In above formulas, $\mathcal{F}$ stands for the amplitudes from the factorizable emission diagrams, and $\mathcal{M}$ for the nonfactorizable ones. In diagrams (a) and (b) of Fig.~\ref{feynman}, when the hard gluons are from the spots ``1" and ``2", their amplitudes are summed into $\mathcal{F}$, and the amplitudes $\mathcal{M}$ is the sum of contributions from spots ``3" and ``4".  $\mathcal{A}$ and $\mathcal{W}$ are the contributions from the annihilation type amplitudes, which are associated with the $W$ boson annihilation and $W$ exchange type process respectively. In practice, in diagrams (c) and (d), the sum of contributions that the gluon are from ``5" and ``6" are denoted as $\mathcal{A}$, and  the contributions that the gluon are from ``7" and ``8" are denoted as $\mathcal{W}$  The superscripts $LL$, $LR$, and $SP$ refer to the contributions from $(V-A)\otimes(V-A)$, $(V-A)\otimes(V+A)$, and $(S-P)\otimes(S+P)$ operators. The subscripts in $\mathcal{F}$ and $\mathcal{M}$ denote the recoiled particles or particle pairs, while the symbols $\mathcal{A}_{a,b}$ and $\mathcal{W}_{a,b}$ means that $a$ and $b$ are the upper and lower particle or particle-pair, as indicated in Fig.~\ref{feynman}. Due to the limitation of space, we will not present their explicit expressions here, which have been given in ref.\cite{Zou:2020atb}. In fact, not all terms can contribute to the decay modes we discussed. In the decays with the $S$-wave $K_0^*(1430)$ as the resonance, the $\mathcal{F}_{K\pi}^{SP}$ will contribute to the decay amplitudes, which disappear in these decays with vector resonance, as the vector structure can not be produced through $(S\pm P)$ currents. Likewise, the $\mathcal{F}_{K\pi}$ will disappear in the decays with the resonance $K_2^*(1430)$ due to the fact that the tensor structure cannot be produced through $V\pm A$ and $S\pm P$ currents.

Last, we can obtain the differential branching fraction
\begin{eqnarray}
\frac{d^2\mathcal{B}}{d \zeta d\omega}=\frac{\tau\omega|\vec{p}_1||\vec{p}_3|}{32\pi^3 m_{B_s}^3}|\mathcal{A}|^2.
 \end{eqnarray}
The magnitudes of three-momenta of the kaon and the bachelor particle in the rest reference frame of the $K\pi$-pair are given by
\begin{eqnarray}
|\vec{p}_1|=\frac{\sqrt{\lambda(\omega^2,m_K^2,m_{\pi}^2)}}{2\omega}, \quad
|\vec{p}_3|=\frac{\sqrt{\lambda(m_{B_s}^2,m_K^2,\omega^2)}}{2\omega},
\end{eqnarray}
with the standard K$\ddot{a}$ll$\acute{e}$n function $\lambda (a,b,c)= a^2+b^2+c^2-2(ab+ac+bc)$.

\section{Numerical Results and Discussions}\label{sec:result}
To perform the theoretical predictions, we should adopt the proper parameters, such as the QCD scale, the mass, lifetime and decay constant of the $B_{s}$ meson, the masses and the widths of the intermediate resonant mesons, and the CKM matrix elements are summarized as follows \cite{Tanabashi:2018oca}:
\begin{gather}
\Lambda_{QCD}^{f=4}=0.25\pm0.05\,{\rm GeV},\;\;m_{B}=5.366\,{\rm GeV},\;\;f_B=0.23\pm 0.02\,{\rm GeV},\;\;\tau_{B_s}=1.509\,ps,\nonumber\\
m_{K^{*\pm}(892)/K^{*0}(892)}=0.892/0.895\,{\rm GeV},\;\;\Gamma_{K^{*\pm}(892)/K^{*0}(892)}=0.0508/0.0474\,{\rm GeV},\nonumber\\
m_{K_2^{*\pm}(1430)/K_2^{*0}(1430)}=1.425/1.432\,{\rm GeV},\;\;\Gamma_{K_2^{*\pm}(1430)/K_2^{*0}(1430)}=0.0985/0.109\,{\rm GeV},\nonumber\\
m_{K_0^{*}(1430)}=1.425\,{\rm GeV},\;\;\Gamma_{K_0^{*}(1430)}=0.270\,{\rm GeV},\;\; |V_{tb}|=1,\;\;|V_{ts}|=0.041,\nonumber\\
|V_{ub}|=(3.65\pm0.12)\times 10^{-3},\;\;|V_{us}|=0.224,\;\;\gamma=(73.5^{+4.2}_{-5.1})^\circ.
\end{gather}

Before presenting our numerical results of the branching fractions and the $CP$ asymmetries, we first declare the theoretical uncertainties considered in this work associated with the nonpertubative parameters, the higher order and power corrections, as well as the CKM matrix elements. In dealing with the hadronic decays of $B$ mesons, the first and foremost uncertainties are from the parameters of the wave functions of the initial and final states, such as the shape parameter $\omega_B=0.5\pm0.05$ GeV and the decay constant $f_B$ in $B_s$ meson wave function, and the Gegenbauer moments in DAs of $K\pi$-pair with different intermediate resonances and also in DAs of the light mesons, which are supposed to be varied with a $20\%$ range in this work. With the improvements of the experiments and the deeper theoretical developments, this kind of uncertainties will be reduced. The second uncertainties are induced by the higher order QCD radiative corrections \cite{Cheng:2014fwa, Cheng:2014gba, Wang:2014mua, Li:2010nn, Li:2012nk} and high power corrections \cite{Shen:2019vdc,Shen:2018abs} of PQCD, which are reflected by varying the $\Lambda_{QCD}=0.25\pm 0.05$ and factorization scale $t$ from $0.8t$ to $1.2t$ for convenience. It is found that this kind of uncertainties in $CP$ asymmetries is comparable with the first one, because the radiative correction and power correction affect the strong phase remarkably. The last errors are the uncertainties of the CKM matrix elements and the CKM angles, which affect the $CP$ asymmetries significantly and have marginal effects on the branching fractions.

\begin{table}[!htb]
\begin{center}
\caption{The flavour-averaged branching ratios (in $10^{-6}$) of $B\to KK\pi$ decays with resonances $K^*(892)$, $K_0^*(1430)$ and $K_2^*(1430)$ in PQCD approach, together with the experimental data \cite{Aaij:2019nmr}.} \label{br}
\begin{threeparttable}
\begin{tabular}{l c c}
 \hline \hline
 \multicolumn{1}{c}{Decay Modes}&\multicolumn{1}{c}{PQCD } &\multicolumn{1}{c}{EXP} \\
\hline\hline
 $B_s \to K^{\pm}(K^{*\pm}(892)\to) \KorKbarz \pi^{\pm}$
&$12.2^{+3.8+4.7+0.5}_{-2.3-2.8-0.0}$
&$12.4\pm0.8\pm0.5\pm2.7\pm1.3$\\
 $B_s \to K^{\pm}(\KorKbarz \pi^{\pm})^*_0$ \tnote{1}
&$26.0^{+11.7+7.9+1.0}_{-9.2-6.1-0.8}$
&$24.9\pm1.8\pm0.5\pm20.0\pm2.6$ \\
 $B_s \to K^{\pm}(K_0^{*\pm}(1430)\to) \KorKbarz  \pi^{\pm}$
&$20.9^{+8.8+6.9+0.7}_{-7.2-4.7-0.2}$
&$19.4\pm1.4\pm0.4\pm15.6\pm2.0\pm0.3$  \\
 $B_s \to K^{\pm}(\KorKbarz \pi^{\pm})_{NR}$
&$13.1^{+6.0+3.7+0.4}_{-5.0-2.9-0.5}$
&$11.4\pm0.8\pm0.2\pm9.2\pm1.2\pm0.5$  \\
 $B_s \to K^{\pm}(K_2^{*\pm}(1430)\to)\KorKbarz \pi^{\pm}$
&$4.4^{+2.2 +1.7+0.5}_{-1.5-1.1-0.0}$
&$3.4\pm0.8\pm0.8\pm5.4\pm0.4$   \\
\hline
 $B_s \to \KorKbarz(\KorKbar\!^{*0}(892)\to) K^{\mp} \pi^{\pm}$
&$11.4^{+4.1+5.1+1.3}_{-1.6-3.0-0.0}$
&$13.2\pm1.9\pm0.8\pm2.9\pm1.4$\\
 $B_s \to \KorKbarz(K^{\mp} \pi^{\pm})^*_0$
&$25.8^{+11.8+8.1+1.1}_{-9.1-5.8-0.1}$
&$26.2\pm2.0\pm0.7\pm7.3\pm2.8$ \\
 $B_s \to \KorKbarz(\KorKbar\!_0^*(1430)\to) K^{\mp}  \pi^{\pm}$
&$20.1^{+8.9+7.0+1.0}_{-6.9-4.4-0.1}$
&$20.5\pm1.6\pm0.6\pm5.7\pm2.2\pm0.3$  \\
 $B_s \to \KorKbarz(K^{\mp} \pi^{\pm})_{NR}$
&$12.9^{+5.9+3.7+0.4}_{-4.9-2.9-0.5}$
&$12.1\pm0.9\pm0.3\pm3.3\pm1.3\pm0.5$  \\
 $B_s \to \KorKbarz(\KorKbar\!_2^*(1430)\to)K^{\mp} \pi^{\pm}$
&$3.4^{+1.6+1.4+0.3}_{-1.1-0.9-0.0}$
&$5.6\pm1.5\pm0.6\pm7.0\pm0.6$   \\
 \hline \hline
\end{tabular}
\begin{tablenotes}
\item $\tnote{1}$ the notation $(K\pi)^*_0$ indicates the total $K\pi$ S-wave modeled by the LASS line shape.
\end{tablenotes}
\end{threeparttable}
 \end{center}
\end{table}

In Table.~\ref{br}, we present our numerical results of the branching fractions with the uncertainties discussed above. In 2019, the LHCb collaboration have achieved their first untagged decay-time-integrated amplitude analysis of $B_s \to K_S^0 K^{\pm}\pi^{\mp}$ decays with the resonances $K^*(892)$, $K_0^*(1430)$, and $K_2^*(1430)$, using a sample corresponding to 3.0 $fb^{-1}$ of $pp$ collision data \cite{Aaij:2019nmr}, and reported the branching fractions with respect to the corresponding resonances, which are also listed for comparison. It is found that within the uncertainties our predictions are in good agreement with the experimental data.

In ref.\cite{Li:2018qrm}, the authors had evaluated the $P$-wave contributions in decays $B_s \to (K\pi) K$ with the branching fractions as
\begin{eqnarray}\label{Yaresults}
\mathcal{B}(B_s\to K^+(K^{*-}\to)K\pi)&=&(7.27^{+1.55+0.45+0.81}_{-1.66-0.37-0.77})\times 10^{-6},\nonumber\\
\mathcal{B}(B_s\to K^-(K^{*+}\to)K\pi)&=&(6.96^{+2.27+1.64+0.32}_{-1.60-1.10-0.31})\times 10^{-6},\nonumber\\
\mathcal{B}(B_s\to K^0(\overline{K}^{*0}\to)K\pi)&=&(6.19^{+1.45+0.12+0.81}_{-1.56-0.14-0.77})\times 10^{-6},\nonumber\\
\mathcal{B}(B_s\to \overline{K}^0(K^{*0}\to)K\pi)&=&(7.16^{+2.55+1.78+0.31}_{-1.84-1.17-0.28})\times 10^{-6}.
\end{eqnarray}
Based on the isospin conservation and the relations
\begin{eqnarray}
\frac{\Gamma(K^{*0}\to K^+\pi^-)}{\Gamma(K^{*0}\to K \pi)}=\frac{2}{3},\,\,\,
\frac{\Gamma(K^{*+}\to K^+\pi^0)}{\Gamma(K^{*+}\to K \pi)}=\frac{1}{3},
\end{eqnarray}
one can obtain the results as
\begin{eqnarray}
\mathcal{B}(B_s \to K^{\pm}(K^{*\pm}(892)\to) \KorKbar \pi^{\pm})
&=& (9.45^{+2.23}_{-1.81})\times 10^{-6},\\
\mathcal{B}(B_s \to \KorKbar^0(\KorKbar\!^{*0}(892)\to) K^{\mp} \pi^{\pm})
&=& (8.90^{+2.36}_{-1.87})\times 10^{-6}.
\end{eqnarray}
One could find that although the above predictions basically match the current LHCb measurements within the errors, the center values are still a bit smaller than our predictions and the currently available values of LHCb collaboration. The discrepancy between the two PQCD predictions originates mainly from the Gegenbauer moments in the DAs of $K\pi$-pair. Because the width of $K^{*}$ is narrow enough, the narrow-width approximation works well here. Under this approximation, the quasi-two-body decay with the resonance $R$ can be factorized as
\begin{eqnarray}
\mathcal{B}(B\to R P\to P_1P_2 P)=\mathcal{B}(B\to R P)\times\mathcal{B}(R\to P_1P_2).
\label{nwa}
\end{eqnarray}
Assuming $\mathcal{B}(K^*(892)\to K\pi)=100\%$ \cite{Tanabashi:2018oca}, we then estimate the branching fractions of two body $B_s \to K^{\pm}K^{*\mp}$ decay and $B_s \to \KorKbar^0 \KorKbar\!^{*0}$ to be
\begin{eqnarray}
\mathcal{B}(B_s \to K^{\pm}K^{*\mp})&=&(18.3^{+9.0}_{-5.4})\times10^{-6},\\
\mathcal{B}(B_s \to \KorKbar^0 \KorKbar\!^{*0} )&=&(17.1_{-5.2}^{+9.9}) \times 10^{-6}.
\end{eqnarray}
In past few years, these two decays have been studied extensively in different theoretical approaches such as the QCD factorization (QCDF)\cite{Cheng:2009mu}, the soft-collinear effective theory (SCET) \cite{Wang:2008rk}, the framework of flavor symmetry \cite{Cheng:2014rfa}, and the PQCD approach \cite{Ali:2007ff}. For comparison, the branching fractions predicted in different approaches are summarized in Table.~\ref{br2}, together with the latest experimental results \cite{Aaij:2019nmr}. It is obvious that our predictions agree well with not only the current LHCb measurements but also other theoretical results. We also noted that the results in \cite{Ali:2007ff} based on PQCD approach are smaller than both our results and the others, which can be improved by adopting the latest DAs of light mesons and keeping the power suppressed terms that are proportional to $(m_K^*/m_B)^2$ especially in the denominator of the quark propagator, as indicated in ref.~\cite{Zou:2015iwa}.

\begin{table}[!htb]
\begin{center}
\caption{The flavour-averaged branching ratios (in $10^{-6}$) of $B\to KK\pi$ decays with resonances $K^*(892)$, $K_0^*(1430)$ and $K_2^*(1430)$ in PQCD approach, together with the experimental data \cite{Aaij:2019nmr}.} \label{br2}
\begin{tabular}{c c c }
\hline\hline
Approach
&$B_s \to K^{\pm}K^{*\mp}$
&$B_s \to \KorKbar^0 \KorKbar\!^{*0}$\\
\hline
 QCDF \cite{Cheng:2009mu}
& $21.6^{+12.1}_{-7.8}$
& $20.6^{+12.4}_{-8.0}$\\\hline
 SCET \cite{Wang:2008rk}
& $19.7^{+5.3}_{-4.5}$
& $18.7^{+5.2}_{-4.4}$\\\hline
 Flavor Symmetry \cite{Cheng:2014rfa}
& $16.01\pm0.91$
& $15.65\pm 0.87$\\\hline
 PQCD  \cite{Ali:2007ff}
& $10.7^{+3.7}_{-2.5}$
& $11.6^{+4.0}_{-2.6}$\\\hline
 This Work
& $18.3^{+9.0}_{-5.4}$
& $17.1_{-5.2}^{+9.9}$\\\hline
 Exp \cite{Aaij:2019nmr}
& $18.6\pm4.7$
& $19.8\pm5.7$\\
\hline
\hline
\end{tabular}
\end{center}
\end{table}

Now, we shall discuss the contribution of $S$-wave $K\pi$-pair, which is related to the resonance $K_0^*(1430)$. Due to the large  interference between the resonant and nonresonant contribuions, the so-called LASS line shape is developed to describe the combined $S$-wave $K\pi$-pair around $1.4$ GeV. As shown in eq.~(\ref{lass}), the first term in the LASS line shape represents the nonresonant contribution while the second one corresponds to the resonant amplitude. In view of this,  we could calculate three type branching fractions and $CP$-violation observables, corresponding to the nonresonant, resonant and the total $S$-wave $K\pi$ contributions. All results are listed in the Table.~\ref{br}, where we can find that our results are in good agreement with the LHCb measurements within errors. In ref.\cite{Wang:2020saq}, the authors studied the contributions of $S$-wave $K\pi$ resonant in the three-body decays $B/B_s^0\to  KK\pi$, where the Gegenbauer moments they used are same as ones of DAs of $K_0^*(1430)$ \cite{Cheng:2013fba}. In fact, the Gegenbauer moments $B_1$ and $B_3$ for the two-meson DAs need not to be identical to that for $K_0^*(1430)$, because they are different nonperturbative quantities describing different objects. As aforementioned, RBW model fails in describing the $S$-wave $K\pi$ resonant contribution around $1.4$ GeV, due to the large interference between resonant and nonresonant contributions.

Using the available branching fraction $\mathcal{B}(K^*_0(1430)\to K\pi)=(93\pm10)\%$ \cite{Tanabashi:2018oca}, we can naively determine the branching fractions of the corresponding two-body decays $B_s \to K^{\pm}K_0^{*\pm}(1430)$ and $B_s \to \KorKbarz \KorKbar\!^{*}_0(1430)$ as
\begin{eqnarray}
&&\mathcal{B}(B_s \to K^{\pm}K_0^{*\pm}(1430))=(33.7_{-15.8}^{+24.3})\times10^{-6},\\
&&\mathcal{B}(B_s \to \KorKbarz \KorKbar\!^{*}_0(1430))=(32.2_{-14.8}^{+24.5})\times10^{-6},
\end{eqnarray}
which are consistent with the measured values from LHCb collaboration
\begin{eqnarray}
&&\mathcal{B}(B_s \to K^{\pm}K_0^{*\pm}(1430))=(31.3\pm2.3\pm0.7\pm25.1\pm3.3)\times10^{-6},\\
&&\mathcal{B}(B_s \to \KorKbarz \KorKbar\!^{*}_0(1430))=(33.0\pm2.5\pm0.9\pm9.1\pm3.5)\times10^{-6}.
\end{eqnarray}
Since the LASS line shape allows us to obtain separately the branching fractions of the contributing parts with respect to resonant part, the effective range part and the coherent sum, after analyzing the predictions of both two decay processes in Table.~\ref{br}, we find that the $K_0^*(1430)$ resonance accounts for about $78\%$, and the effective range shares as large as $46\%$, which implies that the destructive interference between the two parts reaches $24\%$. The same conclusion has been also drawn in ref.\cite{Aaij:2019nmr}.

Now, we move to analyze the decays with the resonance $K_2^*(1430)$. We note that in these decays the $D$-wave $K\pi$-pair cannot be emitted and only be recoiled, due to the fact that the tensor structure can not be produced through the $(V\pm A)$ and $(S\pm P)$ currents. Therefore, this type of quasi-two-body decays has the small branching fractions in comparison with those decays with the $S$ and $P$ waves. The theoretical results are presented in Table.~\ref{br}, which are basically in accordance with the LHCb measurements within the errors. From the table, one can find that the branching fraction of the $B_s \to K^{\pm}(K_2^{*\pm}(1430)\to)\KorKbarz \pi^{\pm}$ is larger than that of $B_s \to \KorKbarz(\KorKbar\!_2^{*0} (1430)\to)K^{\mp} \pi^{\pm}$, because the former decay process gets the enhancement from the colour-allowed tree level emission diagrams with the $K$ meson emitted, while the latter is a pure penguin process.  Again, using the narrow-width approximation and the branching fraction $\mathcal{B}(K_2^*(1430)\to K\pi)=(49.9\pm 1.2)\%$, we can determine the branching fractions of the associated two-body decays as
\begin{eqnarray}
\mathcal{B}(B_s \to K^{\pm}K_2^{*\pm}(1430))&=&(13.2_{-5.8}^{+8.9})\times10^{-6},\\
\mathcal{B}(B_s \to \KorKbarz \KorKbar\!_2^{*0} (1430))&=&(10.2_{-4.3}^{+7.0})\times10^{-6},
\end{eqnarray}
which are consistent with the experimental data
\begin{eqnarray}
\mathcal{B}(B_s \to K^{\pm}K_2^{*\pm}(1430))&=&(10.3\pm2.5\pm1.1\pm16.3\pm1.1)\times10^{-6},\\
\mathcal{B}(B_s \to \KorKbarz \KorKbar\!_2^{*0} (1430))&=&(16.8\pm4.5\pm1.7\pm21.2\pm1.8)\times10^{-6}.
\end{eqnarray}
We also note that for the central values of the $B_s \to \KorKbarz \KorKbar\!_2^{*0} (1430)$ there exists discrepancy between our predictions and experimental data, and it is acceptable because the uncertainties in both sides are rather large. So, the theoretical calculation and experimental measurements with high precision in future are needed. Furthermore, our current predictions are basically in agreement with the previous studies \cite{Qin:2014xta} based on PQCD approach within errors.

In  the light of the isospin symmetry, we obtain the relation between the decays $B_s\to K^{\mp}(K^{*\pm}\to )K^{\pm}\pi^0$, $B_s \to \KorKbarz(\KorKbar\!^{*0}\to) \KorKbarz\pi^0$ and the decays we concerned as follows
\begin{eqnarray}
R=\frac{B_s\to K^{\mp}(K^{*\pm}\to) K^{\pm}\pi^0}{B_s \to K^{\pm}(K^{*\pm}\to) \KorKbarz \pi^{\pm}}=\frac{B_s \to \KorKbarz(\KorKbar\!^{*0}\to) \KorKbarz\pi^0}{B_s \to \KorKbarz(\KorKbar\!^{*0}\to) K^{\mp} \pi^{\pm}}=\frac{1}{2},
\end{eqnarray}
which can also be confirmed by the narrow-width approximation, since the branching ratios of $\KorKbar\!^{*0} \to K^{\pm}\pi^{\mp}$ and $K^{*\pm}\to \KorKbarz \pi^{\pm}$ are the two times larger than the corresponding processes $\KorKbar\!^{*0}  \to \KorKbarz\pi^{0}$ and $K^{*\pm}\to K^{\pm} \pi^0$, respectively.

As is known to all, about 50 years ago the phenomenon of $CP$ violation was discovered and headed to interpret the imbalance between matter and anti-matter. Therefore, it has always being the hot topic in heavy flavor physics and attracted a lot of attention. In SM, the CKM mechanism involving a complex parameter provides the weak phases to satisfy the requirement of $CP$ asymmetry. However, the CKM mechanism for producing $CP$ violation was found to be several orders of magnitude too small to explain the matter domination in the Universe. Thus both experimentalists and theorists have been on the lookout for sources of $CP$ violation beyond SM, and such searching is also one of motivations for searching for new physics beyond SM. Compared with the $B_{u,d}$ system, some $B_s$ decays offers an excellent opportunity to probe the effects of new physics, because in SM the $CP$ violation effects are suppressed and are expected to be small. For example, the angle $\beta$ describing the mixing of $B_d$ system is proved to be of order of $22^{\circ}$, while the mixing angle $\beta_s$ in $B_s$ system is as tiny as $1^{\circ}$, which may increase the new physics sensitivity with more accurate measurements. In addition, compared to the two-body $B$ decays, the multibody decays exhibit much larger $CP$ asymmetries in various regions of phase space, which  would be useful for exploring the abundant sources of $CP$ violation both at low and high invariant mass. Overall, the full QCD-based theoretical analysis of these decays is still missing and model dependent. In ref.\cite{Mannel:2020abt}, the authors introduce a model ansatz to uncover the mechanism of $CP$ asymmetries and emphasize the importance of the open-charm threshold in the high invariant mass region. In refs.\cite{Cheng:2013dua, Cheng:2014uga}, the authors have also analyzed the direct $CP$ violations in charmless three-body decays of $B/B_s$ decays in detail within a simple model based on the framework of the factorization approach. Since the sources of the $CP$ violation are so complicated and not well established clearly, we firstly study the resonant contributions which can be evaluated by adopting proper models.

Motivated above discussions, we will take the decay ${\overline B}_s^0 /{ B}_s^0\to (\KorKbarz\pi^\pm)K^\mp$ as an example and study the resonant contributions to the $CP$-violation observables of these considered three-body decays. As a neutral meson, the flavour eigenstate $B_s^0$ can transform into its anti-particle $\overline{B}_s^0$ via box diagrams, so at the time $t$ the $|B_s(t)\rangle$ produced from the $B_s^0$ at $t=0$ will also have components of $B_s^0$ and $\overline{B}_s^0$. As a result, the $CP$ asymmetries of them are very complicated. Here one studies the four time-dependent decay widths for ${\overline B}_s^0 /{ B}_s^0\to (\KorKbarz\pi^\pm)K^\mp$ decays, the widths of which can be written as
\begin{eqnarray}
{\cal A}_f=|B_s^0\to f\rangle=\langle (K^0\pi^+)K^-|H_{eff}|B_s^0\rangle,\,\,\,\,
\overline{ \cal A}_f=|\overline B_s^0\to f\rangle=\langle (K^0\pi^+)K^-|H_{eff}|\overline B_s^0\rangle,\nonumber\\
\overline{ \cal A}_{\bar f}=|\overline B_s^0\to \bar f\rangle=\langle (\overline K^0\pi^-)K^+|H_{eff}|\overline B_s^0\rangle,\,\,\,\,
{\cal A}_{\bar f}=|B_s^0\to \bar f\rangle=\langle (\overline K^0\pi^-)K^+|H_{eff}|B_s^0\rangle,
\end{eqnarray}
The matrix elements of $|\overline B_s^0\to f\rangle$ and $|\overline B_s^0\to \bar f\rangle$ have been given in eqs.~(\ref{af}) and (\ref{afbar}), respectively. The matrix elements $|B_s^0\to f\rangle$ and $| B_s^0\to \bar f\rangle$ are obtained by changing the signs of the weak phases contained in the products of the CKM matrix elements. Since the flavor eigenstate $B_s$ can transform into anti-particle $\bar{B}_s$ and the physical eigenstates of the mesons with definite mass and decay rate can be presented as the linear combinations as follows
\begin{eqnarray}
|B_{s}^{L,H}\rangle=p|B_s^0\rangle\pm q|\overline{B}_s^0\rangle,
\end{eqnarray}
with $|p|^2+|q|^2=1$, and
\begin{eqnarray}
\frac{q}{p}=\frac{V^*_{tb}V_{ts}}{V_{tb}V^*_{ts}}=e^{-2i\beta_s}.
\end{eqnarray}
So, $|q/p| = 1$, and this ratio has only a phase given by $-2\beta_s$.  Here we neglect the tiny difference between the mass eigenstates and the $CP$ eigenstates with the $B_s^{L(H)}$ being the CP even (odd) state as suggested in ref. \cite{Lenz:2006hd}.

After considering the time evolution of the decay rate, the three-body decay width $\Gamma$ of $|B_s^0(t)\rangle$ decay to the final state $(K^0\pi^+)K^-$  depends on the time $t$ and invariant mass $\omega$\cite{Blusk:2012it},
\begin{multline}
\Gamma[B_s^0(t)\to f](\omega,t)=\frac{1}{2}|\mathcal{A}_f|^2(1+|\lambda_f|^2)e^{-\Gamma_s t}\bigg[\cosh\Big(\frac{\Delta\Gamma_s t}{2}\Big)
+D_f\sinh\Big(\frac{\Delta\Gamma_s t}{2}\Big) \\
+C_f\cos(\Delta m_s t)-S_f\sin(\Delta m_s t)\bigg],
\end{multline}
\begin{multline}
\Gamma[\overline{B}_s^0(t)\to f](\omega,t)=\frac{1}{2}|\mathcal{A}_f|^2\mid\frac{p}{q}\mid^2(1+|\lambda_f|^2)e^{-\Gamma_s t}\bigg[\cosh\left(\frac{\Delta\Gamma_s t}{2}\right)
+D_f\sinh\left(\frac{\Delta\Gamma_s t}{2}\right) \\
-C_f\cos(\Delta m_s t)+S_f\sin(\Delta m_s t)\bigg],
\end{multline}
\begin{multline}
\Gamma[{\overline B}_s^0(t)\to \bar f](\omega,t)=\frac{1}{2}
|\overline {\mathcal{A}}_{\bar f}|^2
(1+|{\overline \lambda}_{\bar f}|^2)
e^{-\Gamma_s t}\bigg[\cosh\Big(\frac{\Delta\Gamma_s t}{2}\Big)
+D_{\bar f}\sinh\Big(\frac{\Delta\Gamma_s t}{2}\Big) \\
+C_{\bar f}\cos(\Delta m_s t)
-S_{\bar f}\sin(\Delta m_s t)\bigg],
\end{multline}
\begin{multline}
\Gamma[{B}_s^0(t)\to \bar f](\omega,t)=\frac{1}{2}
|\overline {\mathcal{A}}_{\bar f}|^2\mid\frac{q}{p}\mid^2
(1+|{\overline \lambda}_{\bar f}|^2)
e^{-\Gamma_s t}\bigg[\cosh\Big(\frac{\Delta\Gamma_s t}{2}\Big)
+D_{\bar f}\sinh\Big(\frac{\Delta\Gamma_s t}{2}\Big) \\
-C_{\bar f}\cos(\Delta m_s t)
+S_{\bar f}\sin(\Delta m_s t)\bigg],
\end{multline}
with
\begin{eqnarray}
\lambda_f=\frac{q}{p}\frac{\overline{\cal A}_f}{{\cal A}_f},\,\,\,\,\,\,
{\overline \lambda}_{\bar f}=\frac{p}{q}\frac{{\cal A}_{\bar f}}{{\overline {\cal A}}_{\bar f}}.
\end{eqnarray}
The symbols $\Delta \Gamma_s$ and $\Delta m_s$ are the width difference and mass difference respectively. In these evolution equations there are six $CP$ asymmetry observables, and they are defined as
\begin{eqnarray}
C_f=\frac{1-|\lambda_f|^2}{1+|\lambda_f|^2}\,,\;\;\;
D_f=\frac{{\rm Re}(\lambda_f)}{1+|\lambda_f|^2}\,,\;\;\;
S_f=\frac{{\rm Im}(\lambda_f)}{1+|\lambda_f|^2};\label{cpo}\\
C_{\bar{f}}=\frac{1-\mid\overline{\lambda}_{\bar{f}}\mid^2}{1+\mid\overline{\lambda}_{\bar{f}}\mid^2}\,,\;\;\;
S_{\bar{f}}=\frac{2{\rm Im}(\overline{\lambda}_{\bar{f}})}{1+\mid\overline{\lambda}_{\bar{f}}\mid^2}\,\;\;\;
D_{\bar{f}}=\frac{2{\rm Re}(\overline{\lambda}_{\bar{f}})}{1+\mid\overline{\lambda}_{\bar{f}}\mid^2}.\label{cpob}
\end{eqnarray}
As for decays ${\optbar B}\!_s^0 \to (K^\pm\pi^\mp){\KorKbarz}$, we set $f=(K^+\pi^-)\overline{K}^0$ in the following discussions.

In the Table.~\ref{cp}, we list all the PQCD predictions to the six $CP$ asymmetry observables with all uncertainties. The parameters $C_f$ and $C_{\bar{f}}$ reflect another type of direct $CP$ violations, which are different from the traditional direct $CP$ violation. From the table, we find that $C_f$ and $C_{\bar{f}}$ in the decays $B_s \to K^{\pm} (K^{*\pm}(892) \to) \KorKbarz \pi^{\pm}$ are rather large, and it is because the tree level transition $b\to s u \bar u $ contributes to the two decays in different ways. For the decay ${\overline B}_s^0 \to(K^0\pi^+)K^-$, $K^-$ is emitted and ($K^0\pi^+$)-pair is recoiled, while for ${B}_s^0 \to(K^0\pi^+)K^-$, ($K^0\pi^+$)-pair is emitted and $K^-$ is recoiled. This reason is also the key factor to explain the large direct $CP$ asymmetries in the two-body $B_s\to K^{*+}K^-/K^{*-}K^+$ decays, as pointed out in ref.~\cite{Ali:2007ff}. However, the decays ${\optbar B}\!_s^0 \to (K^\pm\pi^\mp){\KorKbarz}$ are pure penguin processes, there are no direct $CP$ asymmetries in these two decays, so we have $C_f=C_{\bar{f}}$, as shown in Table.~\ref{cp}. It is also found that $D_f$ ($D_{\bar{f}}$) and $S_f$ ($S_{\bar{f}}$) are large, which indicates that the four decay amplitudes are comparable in magnitude and interfere strongly. These results could be tested in the ongoing LHCb experiments.
\begin{table}[!htb]
\caption{The $CP$-violation observables in $B\to KK\pi$ decays with resonances $K^*(892)$, $K_0^*(1430)$ and $K_2^*(1430)$ in PQCD approach.}
\label{cp}
\begin{tabular}{l c c c}
 \hline
 \hline
 \multicolumn{1}{c}{Decay Modes}&\multicolumn{1}{c}{$C_f$}&\multicolumn{1}{c}{$D_f$}&\multicolumn{1}{c}{$S_f$} \\
\hline\hline
 $B_s \to K^{\pm}(K^{*\pm}(892)\to)\KorKbarz \pi^{\pm}$
&$-0.54^{+0.05+0.07+0.00}_{-0.30-0.16-0.12}$
&$-0.61_{-0.10-0.04-0.00}^{+0.35+0.07+0.08}$
&$-0.57_{-0.15-0.02-0.03}^{+0.20+0.12+0.10}$\\
 $B_s \to K^{\pm}(\KorKbarz \pi^{\pm})^*_0$
&$-0.03^{+0.03+0.04+0.00}_{-0.11-0.14-0.05}$
&$0.93_{-0.04-0.00-0.00}^{+0.05+0.03+0.02}$
&$-0.35_{-0.09-0.00-0.00}^{+0.16+0.10+0.07}$ \\
 $B_s \to K^{\pm}(K_0^{*\pm}(1430)\to) \KorKbarz \pi^{\pm}$
&$-0.01_{-0.10-0.13-0.04}^{+0.07+0.07+0.00}$
&$0.91_{-0.05-0.00-0.00}^{+0.06+0.04+0.03}$
&$-0.39_{-0.09-0.00-0.00}^{+0.17+0.11+0.07}$   \\
 $B_s \to K^{\pm}(\KorKbarz \pi^{\pm})_{NR}$
&$-0.16^{+0.06+0.03+0.00}_{-0.08-0.16-0.05}$
&$0.93_{-0.04-0.01-0.00}^{+0.05+0.01+0.01}$
&$-0.34_{-0.08-0.00-0.00}^{+0.15+0.09+0.06}$  \\
 $B_s \to K^{\pm}(K_2^{*\pm}(1430)\to)\KorKbarz \pi^{\pm}$
&$0.16^{+0.30+0.25+0.20}_{-0.00-0.00-0.00}$
&$0.79_{-0.21-0.33-0.06}^{+0.00+0.00+0.00}$
&$-0.58_{-0.15-0.10-0.00}^{+0.07+0.00+0.07}$  \\
 $B_s \to \KorKbarz(\KstarIzoptbar\to) K^{\mp} \pi^{\pm}$
&$0.061^{+0.09+0.04+0.02}_{-0.20-0.07-0.04}$
&$-0.85_{-0.10-0.05-0.01}^{+0.10+0.00+0.00}$
&$-0.51_{-0.13-0.00-0.00}^{+0.18+0.10+0.03}$\\
 $B_s \to \KorKbarz(K^{\mp} \pi^{\pm})^*_0$
&$-0.02^{+0.19+0.09+0.05}_{-0.02-0.06-0.00}$
&$0.93_{-0.05-0.01-0.01}^{+0.04+0.02+0.00}$
&$-0.37_{-0.11-0.01-0.02}^{+0.12+0.05+0.00}$\\
 $B_s \to \KorKbarz(\KorKbar\!^*_0(1430)\to) K^{\mp}  \pi^{\pm}$
&$0.03^{+0.21+0.07+0.05}_{-0.01-0.07-0.00}$
&$0.90_{-0.08-0.01-0.01}^{+0.06+0.02+0.00}$
&$-0.42_{-0.12-0.01-0.02}^{+0.14+0.08+0.00}$  \\
 $B_s \to \KorKbarz(K^{\mp} \pi^{\pm})_{NR}$
&$-0.13^{+0.06+0.10+0.04}_{-0.05-0.12-0.00}$
&$0.91_{-0.03-0.01-0.00}^{+0.06+0.04+0.01}$
&$-0.38_{-0.08-0.02-0.01}^{+0.15+0.09+0.01}$ \\
 $B_s \to \KorKbarz(\KorKbar\!^*_2(1430)\to)K^{\mp} \pi^{\pm}$
&$-0.21^{+0.34+0.24+0.18}_{-0.00-0.00-0.00}$
&$0.93_{-0.02-0.01-0.00}^{+0.06+0.01+0.03}$
&$-0.29_{-0.05-0.07-0.04}^{+0.03+0.00+0.07}$  \\
\hline
&\multicolumn{1}{c}{$C_{\bar{f}}$}
&\multicolumn{1}{c}{$D_{\bar{f}}$}
&\multicolumn{1}{c}{$S_{\bar{f}}$}\\
\hline
 $B_s \to K^{\pm}(K^{*\pm}(892)\to) \KorKbarz \pi^{\pm}$
&$0.61^{+0.11+0.10+0.05}_{-0.14-0.11-0.06}$
&$-0.74_{-0.12-0.07-0.03}^{+0.11+0.07+0.03}$
&$-0.26_{-0.08-0.11-0.03}^{+0.11+0.17+0.05}$\\
 $B_s \to K^{\pm}(\KorKbarz \pi^{\pm})^*_0$
&$0.08^{+0.09+0.01+0.00}_{-0.10-0.10-0.12}$
&$0.94_{-0.03-0.00-0.00}^{+0.04+0.02+0.01}$
&$-0.31_{-0.06-0.00-0.00}^{+0.14+0.06+0.02}$ \\
 $B_s \to K^{\pm}(K_0^{*\pm}(1430)\to) \KorKbarz  \pi^{\pm}$
&$0.13^{+0.14+0.03+0.00}_{-0.13-0.12-0.13}$
&$0.93_{-0.06-0.00-0.00}^{+0.06+0.02+0.01}$
&$-0.35_{-0.08-0.00-0.02}^{+0.14+0.05+0.02}$ \\
 $B_s \to K^{\pm}(\KorKbarz \pi^{\pm})_{NR}$
&$-0.06^{+0.10+0.06+0.05}_{-0.04-0.13-0.03}$
&$0.95_{-0.03-0.01-0.00}^{+0.03+0.01+0.01}$
&$-0.30_{-0.09-0.01-0.01}^{+0.12+0.05+0.02}$  \\
 $B_s \to K^{\pm}(K_2^{*\pm}(1430)\to)\KorKbarz \pi^{\pm}$
&$-0.42^{+0.34+0.20+0.10}_{-0.03-0.00-0.00}$
&$0.90_{-0.21-0.00-0.00}^{+0.11+0.07+0.03}$
&$-0.09_{-0.12-0.12-0.14}^{+0.00+0.02+0.00}$  \\
 $B_s \to \KorKbarz(\KstarIzoptbar\to) K^{\mp} \pi^{\pm}$
&$0.061^{+0.19+0.11+0.00}_{-0.18-0.10-0.00}$
&$-0.83_{-0.09-0.01+0.00}^{+0.08+0.05-0.00}$
&$-0.54_{-0.09-0.01-0.00}^{+0.17+0.10+0.01}$\\
 $B_s \to \KorKbarz(K^{\mp} \pi^{\pm})^*_0$
&$-0.02^{+0.10+0.09+0.02}_{-0.07-0.06-0.02}$
&$0.94_{-0.08-0.01-0.00}^{+0.02+0.00+0.01}$
&$-0.33_{-0.07-0.01-0.01}^{+0.18+0.01+0.02}$\\
 $B_s \to \KorKbarz(\KorKbar\!^*_0(1430)\to) K^{\mp}  \pi^{\pm}$
&$0.03_{-0.10-0.07-0.03}^{+0.11+0.10+0.00}$
&$0.92_{-0.10-0.00-0.01}^{+0.04+0.00+0.01}$
&$-0.38_{-0.19-0.01-0.01}^{+0.10+0.02+0.01}$ \\
 $B_s \to \KorKbarz(K^{\mp} \pi^{\pm})_{NR}$
&$-0.13^{+0.06+0.12+0.02}_{-0.05-0.10-0.00}$
&$0.93_{-0.06-0.01-0.00}^{+0.04+0.01+0.01}$
&$-0.34_{-0.15-0.01-0.00}^{+0.10+0.04+0.03}$ \\
 $B_s \to \KorKbarz(\KorKbar\!^*_2(1430)\to)K^{\mp} \pi^{\pm}$
&$-0.21^{+0.33+0.25+0.13}_{-0.00-0.00-0.00}$
&$0.95_{-0.02-0.00-0.03}^{+0.03+0.02+0.00}$
&$-0.22_{-0.19-0.07-0.17}^{+0.00+0.01+0.00}$  \\
\hline
\hline
\end{tabular}
\end{table}
\section{Summary}\label{summary}
In this work, motivated by the latest LHCb measurements, we have investigated the quasi-two-body decays $B_s \to \KorKbarz K^{\pm}\pi^{\mp}$ with the $S$, $P$, $D$ partial wave intermediate states $K^*_0(1430)$, $K^*(892)$, and $K_2^*(1430)$ to predict the branching ratios by choosing appropriate $K\pi$ pair wave function within the perturbative QCD approach. The branching fractions we calculated are in good agreement with  experimental results. In previous studies, the decays with $K^*_0(1430)$ and $K^*(892)$ as resonances have been explored with some different wave functions or the model of line shape describing the inner interactions in the $K\pi$-pair. In comparison, both resonant and nonresonant contributions are included in our calculations. Using the narrow-width approximation and the well measured branching fractions of $K^*\to K\pi$, we have also estimated the branching fractions of the two-body decays $B_s \to K K^*$ decays, which are in good agreement with the experimental data and other previous predictions based on QCDF, SCET, PQCD and the flavour symmetry within the uncertainties. Based on the isospin symmetry and the narrow-width approximation, we can get the relationship between the $B_s\to(K^{\pm}\pi^0)K^{\mp}$, $B_s\to(\KorKbarz\pi^0)\KorKbarz$ and the considered decays in this work, and then the branching fractions can be obtained directly. Because the final states are not flavour-specific and both $B_s$ and $\overline{B}_s$ can decay to them with comparable decay amplitudes, the large interference will lead to large $CP$ asymmetries. The six observables have also been calculated, which can be tested in the ongoing LHCb experiment.
\section*{Acknowledgment}
We warmly thank Hsiang-nan Li and Hai-Yang Cheng for constructive suggestions and reading the manuscript carefully. This work is supported in part by the National Science Foundation of China under the Grant Nos.~11705159, 11975195, 11875033 and 11765012, and by the Natural Science Foundation of Shandong province under the Grant No. ZR2018JL001 and No.ZR2019JQ04. X. Liu is also supported by the Qing Lan Project of Jiangsu Province under Grant No. 9212218405, and by the Research Fund of Jiangsu Normal University under Grant No. HB2016004. Zhi-Tian Zou also acknowledges the Institute of Physics Academia Sinica for their hospitalities during the part of the work to be done.
\bibliographystyle{bibstyle}
\bibliography{mybibfile}
\end{document}